# C-Band Lithium Niobate on Silicon Carbide SAW Resonator with Figure-of-Merit of 124 at 6.5 GHz

Tzu-Hsuan Hsu, *Graduate Student Member, IEEE*, Joshua Campbell, *Graduate Student Member, IEEE*, Jack Kramer, *Graduate Student Member, IEEE*, Sinwoo Cho, *Graduate Student Member, IEEE*, Ming-Huang Li, *Senior Member, IEEE*, and Ruochen Lu, *Member, IEEE*

***Abstract*—In this work, we demonstrate a C-band shear-horizontal surface acoustic wave (SH-SAW) resonator with high electromechanical coupling ($k_t^2$) of 22% and a quality factor (*Q*) of 565 based on a thin-film lithium niobate (LN) on silicon carbide (SiC) platform, featuring an excellent figure-of-merit (FoM = $k_t^2 \cdot Q_{max}$) of 124 at 6.5 GHz, the highest FoM reported in this frequency range. The resonator frequency upscaling is achieved through wavelength (λ) reduction and the use of thin aluminum (Al) electrodes. The LN/SiC waveguide and synchronous resonator design collectively enable effective acoustic energy confinement for a high FoM, even when the normalized thickness of LN approaches a scale of 0.5λ to 1λ. To perform a comprehensive study, we also designed and fabricated five additional resonators, expanding the λ studied ranging from 480 to 800 nm, in the same 500 nm-thick transferred Y-cut thin-film LN on SiC. The fabricated SH-SAW resonators, operating from 5 to 8 GHz, experimentally demonstrate a $k_t^2$ from 20.3% to 22.9% and a *Q* from 350 to 575, thereby covering the entire C-band with excellent performance.**

## I. Introduction

THE advancement of fifth-generation (5G) wireless communication, driven by the growing demand for mobile data from online gaming, video streaming, and other data-intensive applications, necessitates the use of larger physical bandwidths [1]. This calls for the development of low-power, wideband, high-performance wireless transceivers, particularly in super high frequency (SHF) spectrums like the C-band (4 – 8 GHz). A significant challenge in this endeavor is the creation of large bandwidth front-end filters with compact designs in SHF range [2]. To address this issue, the emergence of miniaturized acoustic filters featuring suspended piezoelectric plates, enabled by advances in micromachining technology and material science, offers a promising solution. Overcoming the limitations of conventional bulk acoustic wave (BAW) filters, which use aluminum nitride (AlN) resonators and suffer from limited electromechanical coupling capping at about 6% [3], the novel piezoelectric microelectromechanical systems (MEMS) filters employing the aluminum scandium (Sc) nitride (AlScN) [4], lithium niobate (LN) [5]-[7], and lithium tantalate (LT) [8] thin films have shown significant promise for addressing wideband filtering above 5 GHz.

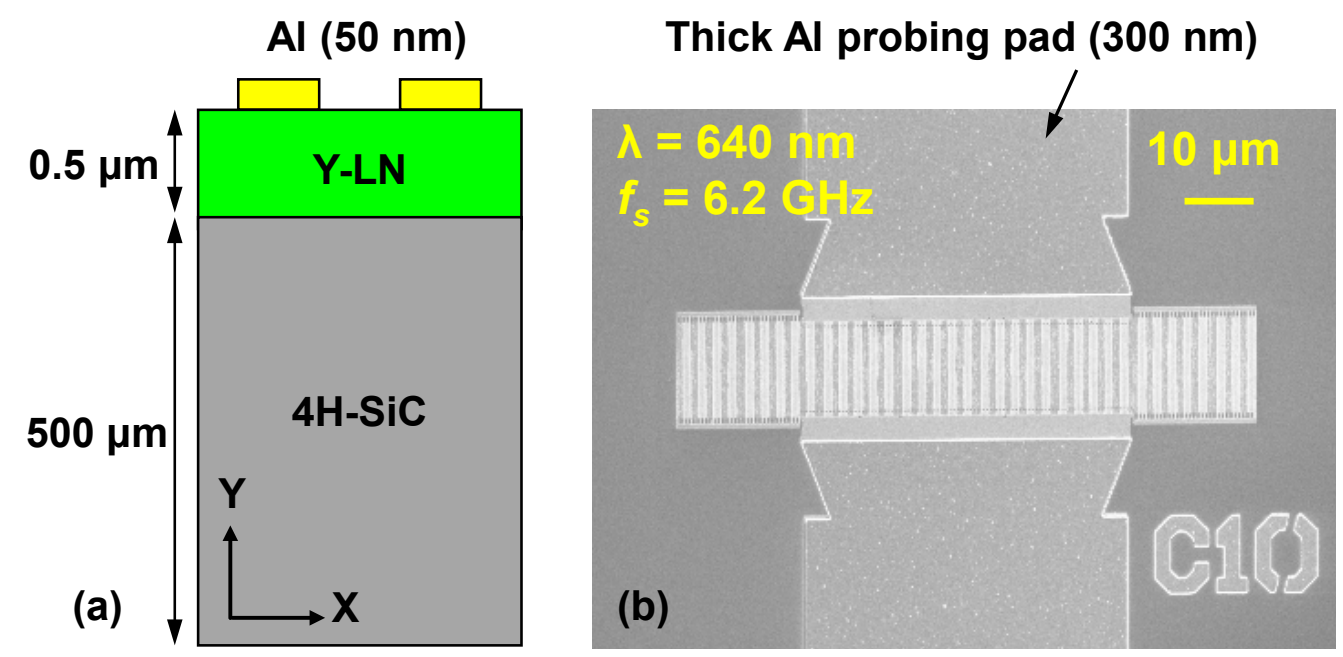

Fig. 1. (a) Unit cell model of the resonator based on LN/SiC. (b) SEM image of the C-band SH-SAW resonator proposed in this study.

The SHF filters based on suspended MEMS piezoelectric resonators exhibit excellent performance, yet they face several challenges. AlScN $S_0$ Lamb wave resonators are promising for multi-band filter implementation on a single chip due to their impressive frequency scalability. However, their electromechanical coupling ($k_t^2$) ranging from 2.88% to 8.07% is limited by the Sc concentration levels and electrode design [9]. In contrast, AlScN BAW resonators potentially offer a higher $k_t^2$ of 13.8% [10] using thickness extensional mode operation compared to Lamb wave resonators, but their complex fabrication process is a significant drawback.

Moreover, $n^{th}$-order Lamb wave resonators utilizing single- and multi-layered transferred LN thin films demonstrate unparalleled performance for filter implementations, with a $k_t^2$ of 31% and a quality factor (*Q*) of 319 at 5 GHz, as well as a $k_t^2$ of 65% and a *Q* of 160 at 17 GHz, as shown in [11]-[13]. Yet, they require individual trimming to achieve precise LN

Manuscript received April 8, 2024. This work was partially supported by DARPA COmpact Front-end Filters at the ElEment-level (COFFEE) Program, partially supported by Sandia's Laboratory Directed Research & Development (LDRD) Program, and partially supported by the National Science and Technology Council (NSTC) of Taiwan under NSTC 112-2221-E-007-103-MY3 (Corresponding author: Ming-Huang Li)

Tzu-Hsuan Hsu is with the Department of Power Mechanical Engineering, National Tsing Hua University, Hsinchu 300044, Taiwan, and also with the Department of Electrical and Computer Engineering, the University of Texas at Austin, TX 78712 USA.

Joshua Campbell, Jack Kramer, Sinwoo Cho and Ruochen Lu are with the Department of Electrical and Computer Engineering, the University of Texas at Austin, Austin, TX 78712 USA.

Ming-Huang Li is with the Department of Power Mechanical Engineering, National Tsing Hua University, Hsinchu 300044, Taiwan (e-mail: mhli@pme.nthu.edu.tw).

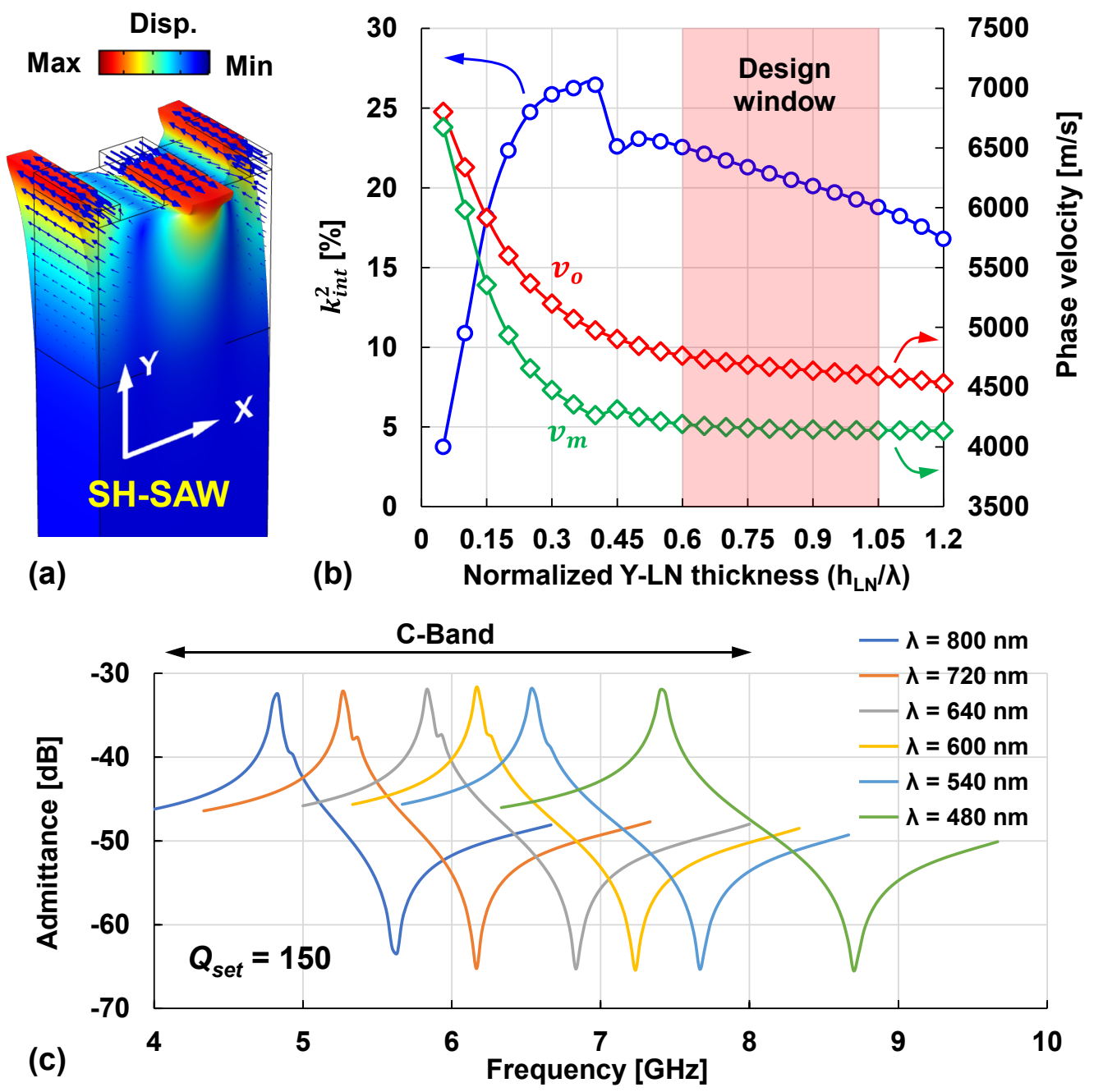

Fig. 2. (a) Simulated mode shape of the SH-SAW resonator. (b) Dispersion behavior of $k_{int}^2$. (c) Simulated admittance spectra of the LN/SiC SH-SAW resonators. The quality factor ($Q_{set}$) in the simulation is 150 (unit cell simulation scaled based on actual designed capacitance).

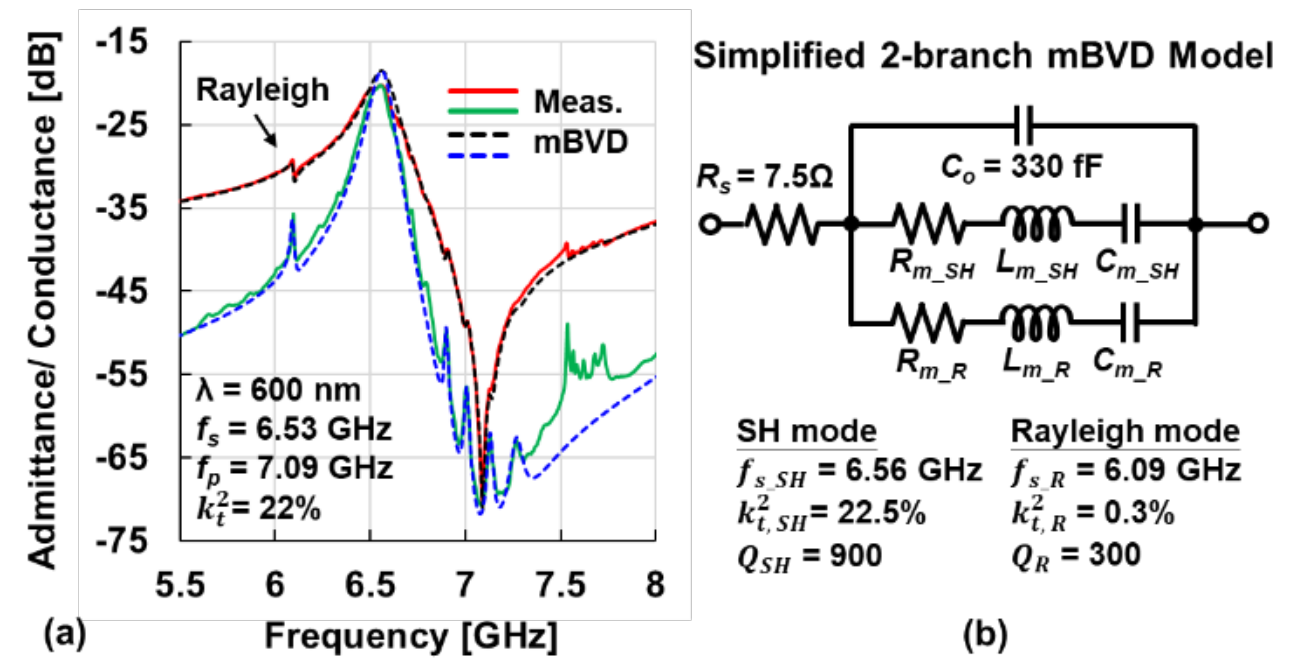

Fig. 3. (a) Measured and mBVD-fitted responses of the C-band SH-SAW resonator with λ = 600 nm. (b) Simplified two-branch mBVD model only considering major modes.

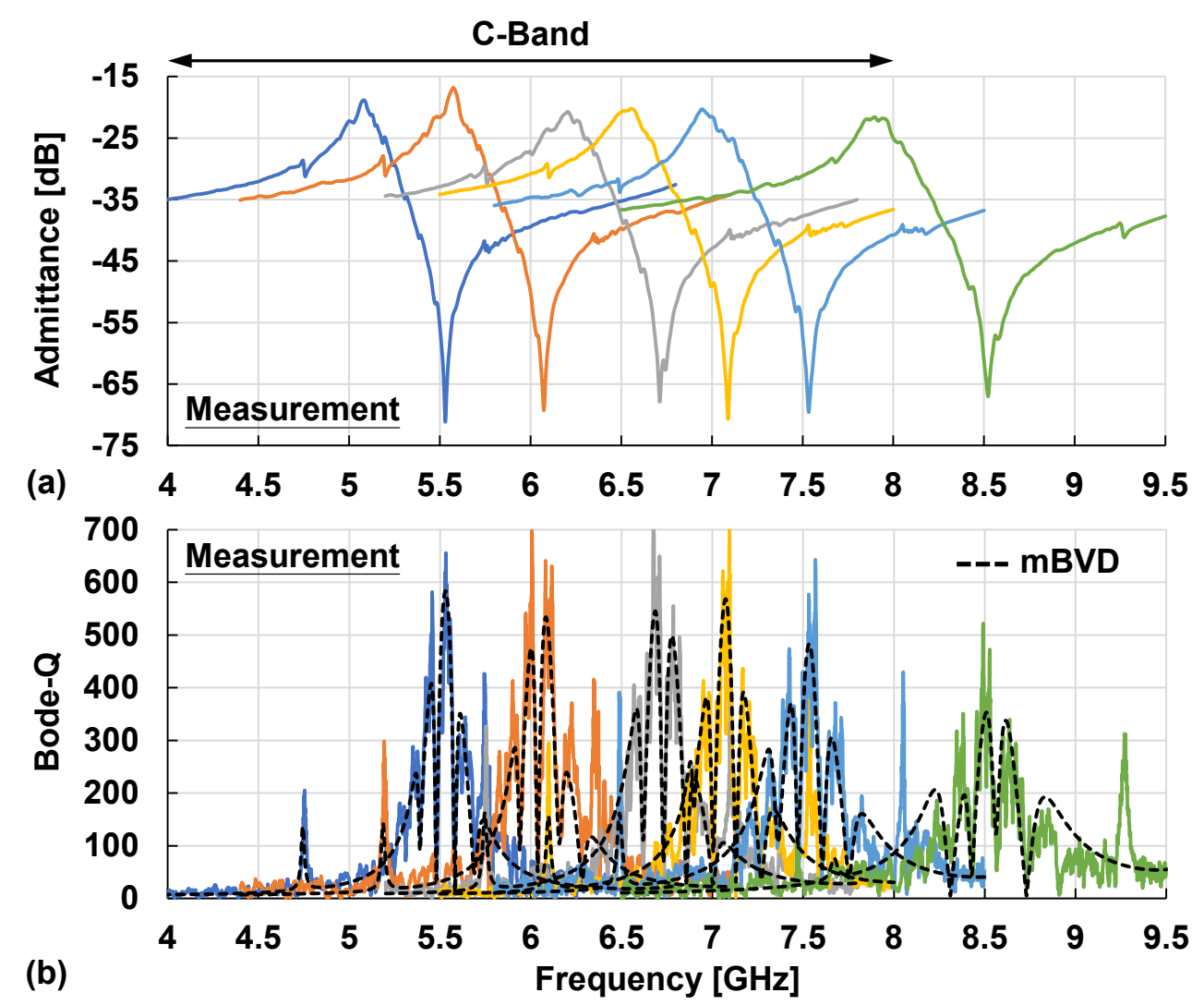

Fig. 4. (a) Measured admittance spectra and (b) extracted Bode-Q of the resonators of the C-band SH-SAW resonators.

thickness for accurate frequency definition and spurious mode elimination, which restricts design flexibility. Finally, despite the exceptional performance of the aforementioned solutions [4]-[13], a potential concern arises from the confinement of acoustic waves in suspended fragile films. This aspect may compromise mechanical robustness in mass production processes.

To address this issue, we present an alternative approach by extending the previously developed thin film surface acoustic wave (TFSAW) technology [14]-[23] to implement C-band shear horizontal (SH) mode resonators. The proposed SH-SAW devices, with a series resonance frequency ranging from 5 GHz to 8 GHz, are designed and characterized on a LN/SiC functional substrate through wavelength scaling. The measured resonators exhibit $k_t^2$ ranging from 20.3 to 22.9%, $Q$ from 350 to 575, and cover the entire C-band from 5.08 to 7.89 GHz, with a high figure of merit of 124 at 6.5 GHz.

## II. Design of C-band SH-SAW Resonators

Fig. 1(a) shows the cross-sectional view of the unit cell model of a LN/SiC SH-SAW resonator, with a LN thickness ($h_{LN}$) of 500 nm. The duty factor of the interdigital electrodes (IDT) is set to 50%, and the excited SH-SAW is directed along the X-axis for optimal $k_t^2$ in Y-cut LN. Fig. 1(b) depicts the scanning electron microscopy (SEM) image of the fabricated C-band resonator, featuring a wavelength (λ) of 640 nm, an aperture (A) of 20λ, 80 IDT electrode pairs ($N_e$), and 60 reflecting gratings ($N_r$). The IDTs are defined by electron beam lithography (EBL) followed by thermal evaporation to deposit a 50 nm-thick aluminum (Al). A second lithography and metal deposition step are taken to form a 300 nm-thick probing pad, which reduces contact resistance during RF measurements.

To evaluate the RF response in the C-band range, a unit cell model is adopted for finite element method (FEM) simulation. The vibration mode shape is depicted in Fig. 2(a), while the intrinsic electromechanical coupling factor ($k_{int}^2$) dispersion is presented in Fig. 2(b). As shown in the displacement distribution profile in Fig. 2(a), the displacement in the vertical direction (i.e., along the Y-axis) quickly attenuates at the LN/SiC surface, effectively confining the acoustic energy to achieve high $Q$. The $k_{int}^2$ is evaluated using Alder's approach [22] without mechanical loading, calculated by $k_{int}^2 = (v_o^2 - v_m^2)/v_o^2$, where $v_o$ and $v_m$ are the phase velocities corresponding to the electrical open surface and electrical short surface of the LN/SiC stack, respectively. The simulations reveal that although $k_{int}^2$ starts to decrease for $h_{LN}$ > 0.4λ, it remains sufficiently large (i.e., $k_{int}^2$ > 20%) to support high-performance C-band resonator designs, with $h_{LN}$ ranging from 0.6λ to 1.05λ in this case. With a 50 nm-thick Al electrode and an IDT duty factor of 50%, Fig. 2(c) exhibits the frequency responses for wavelengths ranging from λ = 800 nm ($h_{LN}$ = 0.625λ) to λ = 480 nm ($h_{LN}$ = 1.04λ). This successfully demonstrates resonators with a coupling factor > 20% within the entire C-band.

TABLE I

COMPARISON OF C-BAND RESONATORS FABRICATED IN THIS WORK

| Device | A | B | C | D | E | F |
|---|---|---|---|---|---|---|
| λ [nm] | 800 | 720 | 640 | 600 | 560 | 480 |
| $h_{LN}/\lambda$ | 0.625 | 0.695 | 0.78 | 0.833 | 0.892 | 1.04 |
| A [λ] | 20 | 20 | 20 | 20 | 20 | 20 |
| $N_e$ | 80 | 80 | 80 | 80 | 80 | 80 |
| $N_r$ | 60 | 60 | 60 | 40 | 40 | 40 |
| $C_o$ [fF] | 450 | 410 | 340 | 330 | 270 | 220 |
| $R_{m_SH}$ [Ω] | 0.443 | 0.42 | 0.47 | 0.443 | 0.51 | 0.53 |
| $f_s$ [GHz] | 5.079 | 5.576 | 6.202 | 6.531 | 6.947 | 7.896 |
| $f_p$ [GHz] | 5.530 | 6.074 | 6.712 | 7.090 | 7.534 | 8.523 |
| $k_t^2$ [%] | 22.9 | 23 | 21.1 | 22 | 21.7 | 20.3 |
| $Q_s$* | 52 | 80 | 38 | 37 | 46 | 40 |
| $Q_{max}$** | 575 | 530 | 540 | 565 | 480 | 350 |
| FoM | 131 | 122 | 114 | 124 | 104 | 71 |

*$Q_s = f_s/f_{BW,3dB}$

**$Q_{max}$ ~ $Q_p$ is extracted from multi-branch mBVD fitting (conservative)

## III. MEASUREMENT RESULTS

The C-band resonators fabricated in this work were characterized with on-wafer RF GSG probing and standard short-open-load-through (SOLT) calibration. The resonator admittance is then derived from the measured S-parameters, which include inherent non-idealities such as parasitic resistance ($R_s$) and capacitance, using PathWave Advanced Design System (ADS) software. No additional de-embedding techniques were performed.

### A. Admittance characterization

Fig. 3(a) illustrates the extracted admittance and conductance responses for a 6.5 GHz resonator, accompanied by accurate fitting results from the multi-branch modified Butterworth-Van Dyke (mBVD) model. A high $k_t^2$ of 22% was directly computed from the series ($f_s$) and parallel resonance frequencies ($f_p$) of the resonator using the formula $k_t^2 \approx (\pi^2/8)\cdot(f_p^2 - f_s^2)/f_s^2$ [24]-[26]. This formula of calculating the $k_t^2$ has been widely adopted by many research groups in the RF acoustic research community for evaluating the resonator performance [24]-[26]. It can also be used with the mBVD model parameters following an approximation as $k_t^2 \approx (\pi^2/8)\cdot(C_m/C_o)$ to link key performance index with modeling parameters.

As a regular IDT design was chosen for this study, several spurious modes arose between $f_s$ and $f_p$, which can be mitigated through IDT apodization in future designs. Some deviations between simulated and measured $f_s$ and $k_t^2$ were noted, potentially caused by uncertain material properties of the ultrathin metal electrode of only 50 nm. Fig. 3(b) shows the fitted $k_t^2$ and $Q$ for the major modes (i.e., SH and Rayleigh waves), with minor modes neglected, highlighting the outstanding performance of the SH mode in terms of $k_{t,SH}^2$ and $Q_{SH}$. Aside from the resistive loss existing in any interdigital transducer design, the large $R_s$ of 7.5Ω in this case can also be attributed to the deposited Al quality such as the hillocks visible in Fig. 1(b) and the relatively thin busline thickness of 350 nm. Future improvements in the deposition recipe and fabrication control are expected to improve the $R_s$ performance.

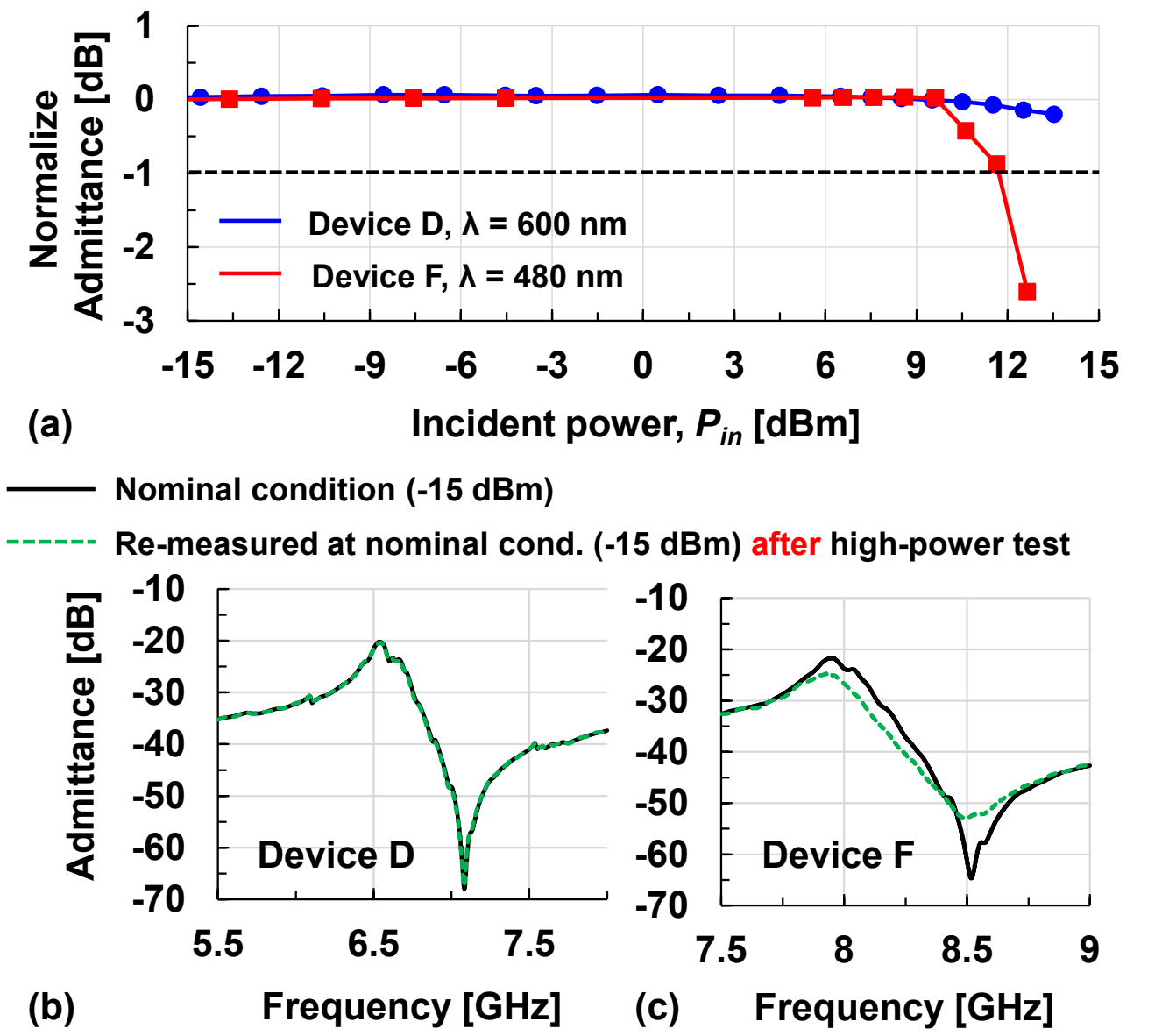

Fig. 5. (a) Normalized admittance for device D and F under different incident power level. (b) Resonator characteristics after high power test where device F shows irreversible device damage.

Figs. 4(a) and (b) show the admittance and Bode-$Q$ [27] of the fabricated C-band resonators from 5 GHz to 8 GHz, respectively. The detailed design parameters and testing results are summarized in Table I. As clearly seen in Table. I, due to the low motional resistance ($R_m$) of the resonator, its admittance ratio and series quality factor ($Q_s$) at $f_s$ is known to highly affected by $R_s$ attributed from electrical interconnect and probing contact. Therefore, to extract the actual performance across the entire frequency spectrum, the Bode-$Q$ method has been adopted and verified experimentally in many prior arts [27]. However, since the measured Bode-$Q$ may exhibit strong fluctuation owing to the existence of spurious modes, it might not be ideal to directly take a number point from the Bode-$Q$ plot as it might provide over-estimated $Q$ value. As a result, in this work, the measured Bode-$Q$ was presented alongside those evaluated using the mBVD model fitting to avoid overclaiming device performance. The maximum Bode-$Q$ ($Q_{max}$) reported is taken from the mBVD model fitted curve. As a result, all resonators designed in this study achieve a large $k_t^2$ > 20% and $Q_{max}$ > 350, yielding FoMs (= $k_t^2 \cdot Q_{max}$) ranging from 71 to 131. Device F, as listed in Table I, exhibits a lower FoM due to a reduced $Q_{max}$ at the high edge of the C-band. As a result, 5 out of the 6 devices reported in this study show a FoM greater than 100, demonstrating significant promise for the proposed approach in RF signal processing.

### B. Power handling capability

To study the power handling capability of the C-band resonators, device D and device F were directly probed at room temperature in air with a frequency-sweep signal applied, and the port impedance was set to 50Ω. The maximum incident RF power in our setup is approximately +13 dBm. As shown in Fig. 5(a), device F (λ = 480 nm) exhibits 1 dB compression characteristics at $P_{in} = P_{1dB}$ = +11.6 dBm. In contrast, device D's $P_{1dB}$ exceeds +13 dBm, despite its similar performance to device

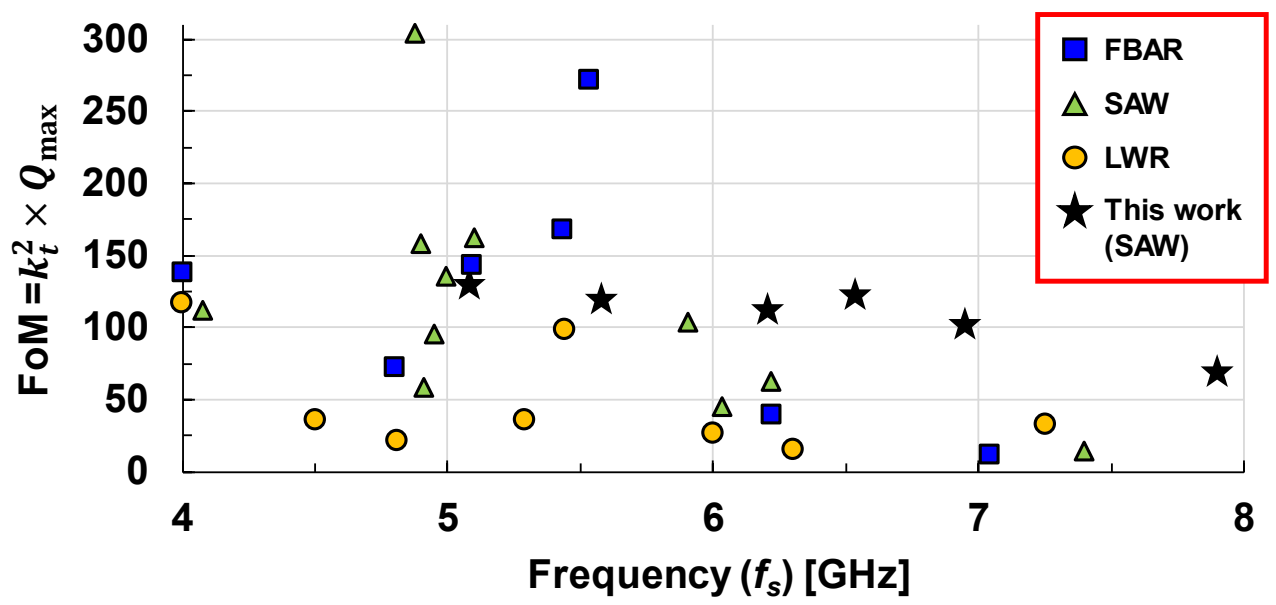

Fig. 6. Surveys of state-of-the-art RF acoustic resonators in C-band. The electromechanical coupling coefficient from different reports is recalculated with $k_t^2 = (\pi^2/8) \cdot (f_p^2 - f_s^2)/f_s^2$ for fair comparison.

F. This is because device D features greater spacing between IDT fingers than device F due to its larger λ (λ = 600 nm), effectively alleviating the effects of acoustomigration and electromigration under high power operation. Furthermore, we re-measured both devices at a low power level of -15 dBm after the high-power test, as shown in Fig. 5(b). However, device F appears to be permanently damaged after the high-power measurement, as indicated by the distorted admittance.

Although our C-band resonator prototype shows limited power handling capability due to fine feature size, it is still comparable to other state-of-the-art acoustic resonators in the same frequency range [9]. To improve power handling in the future, employing composite metals or engineered metal alloys as the electrode material instead of pure Al could be an effective approach [9][35]. Additionally, optimizing the circuit topology for better heat and power distribution may be beneficial for the practical use of the proposed device in filter applications.

### C. Comparison

Finally, Fig. 6 compares the FoM of reported resonators in the C-band, including the Lamb wave [5], [8], [9], [11], [28]-[30], BAW [3], [4], [10], [31]-[33], and SAW resonators [11], [15]-[21] made of AlN, AlScN, LN, or LT, respectively. To ensure a fair comparison, given the variability in definitions of the electromechanical coupling coefficient across different reports, the FoM values presented in Fig. 6 are meticulously calculated based on a unified definition. In general, the devices A to F presented in this work exhibits comparable FoMs with mainstream state-of-the-art devices (SOAs) ranging from 4 to 6 GHz and demonstrates superior FoMs than SOAs at 6 to 8 GHz across different technology platforms, underscoring an excellent potential in frequency scaling of the LN/SiC SH-SAW platform.

## V. Conclusion

We demonstrate LN/SiC SH-SAW resonators with stable performance across most of the spectrum in C-band (5 to 8 GHz range), achieving highest FoM of 124 at 6.5 GHz and an overall high FoM over 100 between 5 to 7 GHz. The adoption of a SiC substrate and the SH-SAW acoustic mode are the key factors in concurrently attaining both high $Q$ and $k_t^2$ in this frequency range. Our results indicate that the LN/SiC SH-SAW resonators can be a critical platform for achieving exceptional FoMs in the C-band.